\documentclass{article}
\usepackage{graphicx}
\usepackage{authblk}
\usepackage[utf8]{inputenc}
\usepackage[style=authoryear]{biblatex}
\usepackage[T1]{fontenc}
\usepackage{enumitem} 
\usepackage{graphicx}
\usepackage{titlesec}
\usepackage{xcolor}
\usepackage{csvsimple}
\usepackage{rotating}
\usepackage{pgfplots}
\usepackage{setspace}
\pgfplotsset{compat=1.18}
\usepackage{caption}
\usepackage{amsmath,amssymb}
\usepackage{hyperref}
\usepackage{array}
\usepackage{booktabs}
\usepackage{pdflscape}
\usepackage{longtable}
\usepackage[margin=1in]{geometry}
\usepackage[
  capitalise,
  nameinlink
]{cleveref}
\usepackage{tikz}
\usetikzlibrary{shapes.geometric, arrows}
\usepackage{standalone}
\usetikzlibrary{calc,fit,positioning}
\tikzset{>=latex}
\usepackage{rotating}
\usepackage{subcaption}
\usepackage{multirow}
\usepackage{varwidth}
\usepackage{tabularx}
\usetikzlibrary {arrows.meta,bending,positioning}
\renewcommand{\arraystretch}{1.1}
\titleformat{\subsubsection}{\itshape}{\thesubsubsection}{1em}{}
\tikzstyle{process} = [rectangle, minimum width=2cm, minimum height=0.7cm, text centered, draw=black, fill=gray!30]
\tikzstyle{actors} = [rectangle, draw, text width=3.3cm, minimum height=1.5cm, text centered, draw=black, fill=gray!30]
\tikzstyle{arrow} = [thick,->,>=stealth]
\tikzstyle{standards} = [rectangle, draw, text width=2cm, text centered, draw=black, fill=gray!5]
\tikzstyle{technology} = [rectangle, draw, text width=2cm, text centered, draw=black, fill=gray!35]
\tikzstyle{users} = [rectangle, draw, text width=2cm, text centered, draw=black, fill=gray!20]
\tikzstyle{dasharrow} = [dashed,thick,->,>=stealth]
\tikzstyle{doubledasharrow} = [dashed,thick,<->,=stealth']
\tikzstyle{doublearrow} = [thick,<->,=stealth']

\title{Irrelevant carrots and non-existent sticks: trust, governance, and security in the transition to quantum-safe systems}
\author[1,2]{Ailsa Robertson}
\author[1]{Siân Brooke}
\author[1,2]{Sebastian De Haro}
\author[1,2]{Christian Schaffner}
\affil[1]{Universiteit van Amsterdam, Amsterdam, Netherlands}
\affil[2]{QuSoft, Amsterdam, Netherlands}
\begin{document}
\maketitle
\begin{abstract}
    Quantum computing poses an urgent and widely recognised threat to global cybersecurity, enabling encrypted government, financial, and healthcare data harvested today to be decrypted in the near future. 
    Transitioning to quantum-safe cryptography is therefore essential, demanding coordinated action across a complex, multi-actor innovation system. 
    Drawing on insights from an expert workshop in Amsterdam, this study develops a socially informed vision for a quantum-safe future and analyses the current innovation landscape to identify critical gaps and the actions needed to address them. 
    We map twelve key actor groups involved in the migration process, finding that regulators exert the strongest direct influence, while standardisation bodies play a crucial indirect role. 
    This research provides one of the first system-level mappings of actors, influence pathways and governance responsibilities shaping the quantum-safe transition, revealing several responsibilities with unclear ownership.
    Although centred on the Netherlands, our findings are applicable to other national contexts navigating quantum-safe transitions.
\end{abstract}
\section{Introduction}
Innovation in quantum computing poses a fundamental challenge to current \textit{cryptographic systems} which use mathematical techniques to secure information, including encryption, authentication, and digital signatures [\cite{pqc_migration_handbook,groenland,timelines,nist2035}].
Everyday applications which use cryptography include communication networks, commerce platforms and digital banking infrastructure [\cite{pqc_migration_handbook}].
\textit{Quantum computers}, which harness the principles of quantum mechanics to enable fundamentally new forms of computation, are fast advancing from theoretical possibility to reality [\cite{timelines,groenland}].
As such technologies evolve, the need to design and implement cryptographic schemes that remain secure in the quantum era has become increasingly urgent: communication, commerce, digital banking and a myriad of other systems must be upgraded in order to resist quantum attacks.
In response, governments, research institutions and private stakeholders are working to develop and adopt cryptographic solutions which are resistant to attacks from quantum computers [\cite{pqc_migration_handbook,nist2035}].

The terminology for this emerging cryptography varies considerably across contexts: standards bodies such as NIST use \textit{post-quantum cryptography}, several industry vendors and policy documents favour \textit{quantum-safe}, while others adopt \textit{quantum-resistant}, often interchangeably [\cite{rfc}]. 
Each term has drawbacks: \textit{post-quantum} can imply that the techniques themselves use quantum properties or are only relevant once large-scale devices exist; conversely, labels such as \textit{safe} or \textit{resistant} may overstate the assurances they offer [\cite{rfc}]. 
We note that different terms offer different framings and, despite its limitations, we adopt the term \textit{quantum-safe cryptography} (QSC) for consistency, reflecting its current prominence across policy, industry, and academic discourse.
Such variation in language highlights the broader challenge this transition poses: coordination across diverse technical, institutional, and policy communities.

While the cryptographic community has made significant advances in algorithmic development and standardisation [\cite{pqc_migration_handbook}], less is known about how institutional actors (such as \textit{regulators, implementers, funders, and industry leaders}) are organising around this transition collectively.
National contexts are critical in this collective effort, as governments play a central role in driving cryptographic change, while also responding to international standards, security concerns, and infrastructural dependencies [\cite{Csenkey01092024,nist2035}].
As with discussions around the policy implications of artificial intelligence, an understanding of the capabilities of quantum computers is important but grasping their technical details is not critical; rather, a working knowledge of how these devices reshape risks and responsibilities is essential for engaging with the national systems and governance structures in which they are embedded [\cite{groenland}].
Hence, while no quantum or technical expertise is required to follow this study, readers interested in the mechanics of quantum computing may refer to Koen Groenland's book \textit{Introduction to Quantum Computing for Business} [\cite{groenland}] for an overview.
\paragraph{Research objectives and contribution.}
This paper investigates the innovation system of quantum-safe cryptography in the Netherlands as a case study in how national systems are responding to the post-quantum imperative. 
Drawing on a half-day participatory mapping workshop held with industry, government and academic experts on January 29, 2025 in Amsterdam, we examine the actors present in this emerging innovation system and the roles and responsibilities they assume. 
We use the term \textit{actors} deliberately to recognise the agency, situated knowledge, and interpretative authority of the organisations involved, not simply as institutional representatives, but as contributors and co-constructors of the innovation system under study [\cite{costanza2020justice,brooke2024freelancers}]. 
This framing allows us to account for how expertise is enacted, how responsibilities are negotiated, and how strategic decisions are shaped from within the innovation system itself. 
By foregrounding the institutional dimensions of cryptographic change, the paper aims to contribute to scholarship on innovation systems, cybersecurity governance, and technological transitions. 
The study is guided by three research questions:
\begin{enumerate}[label=RQ\arabic*]
    \item Who are the key organisational actors involved in the development and deployment of quantum-safe cryptography in the Netherlands? \label{RQ1}
    \item What roles and responsibilities do these actors assume in supporting the transition to quantum-safe cryptography? \label{RQ2}
    \item How can the (Dutch) transition to a quantum-safe innovation system be managed proactively? \label{RQ3}
\end{enumerate}
These research questions address a notable gap in quantum-safe cryptography literature, which has predominantly focused on technical standardisation and associated challenges, with limited attention paid to the socio-institutional landscape in which these developments, for instance see \cite{KONG2024101884,Csenkey01092024,kong_22}.
The scholarly contribution of this study is threefold: empirical, methodological, and application-based.
First, the study advances understanding of how policy coordination failures constrain the effectiveness of the Dutch national innovation system by systematically mapping the emerging innovation system surrounding quantum-safe cryptography in the Netherlands and analysing the roles, responsibilities, and situated interactions of its key actors.  
Second, the study demonstrates the methodological value of a qualitative, iterative focus group design for eliciting stakeholder insights and refining analytical frameworks.
Third, the study demonstrates the application-oriented value of integrating technically informed perspectives into a social scientific study, enabling effective communication and engagement with stakeholders across technical and policy domains.
Framing the cryptographic transition as an innovation-system challenge rather than solely a technological issue, we show how agency, expertise, and institutional coordination collectively shape societal readiness for the quantum threat, allowing us to examine why the system's carrots (incentives) and sticks (penalties) are insufficient to prompt widespread transition.
\section{Theoretical framework}
This study draws on innovation systems theory, organisational dynamics and sociotechnical imaginaries, to frame the Dutch transition to quantum-safe cryptography and anticipated barriers in its management, positioning the migration not merely as a technical shift but as a complex process involving distributed coordination, institutional trust, and contested futures. 
The following subsections outline the theoretical concepts that inform our analysis: the boundaries and dynamics of NISs; responsibilities and trust within organisational dynamics; and the application of sociotechnical imaginaries for transition policy pathways. 
These provide the conceptual grounding for the discussion of our findings in \Cref{sec:discussion}.  
\subsection{\textit{Key terms}}\label{sec:key_terms}
We elaborate on key terms to provide the necessary background to grasp the nature of the quantum threat, as well as for clarity, as meanings can shift across contexts which can cause misalignment even among experts.
A table of acronyms is provided in \cref{tab:acronyms}.
\paragraph{Cryptographic systems.}
Cryptographic systems enable secure communication by ensuring \textit{confidentiality}, \textit{authenticity} and \textit{integrity}. 
These properties are achieved through \emph{encryption} and \emph{digital signatures}. 
Historically, cryptography has been divided into \textit{symmetric} and \textit{asymmetric} methods, with the latter relying on mathematical problems that are computationally difficult for conventional computers [\cite{pqc_migration_handbook,groenland,timelines}].
Analogous to the way a computer’s architecture is built from multiple software components, a cryptographic system is constructed from fundamental elements known as \emph{cryptographic primitives} [\cite{pqc_migration_handbook}].
\paragraph{Quantum properties.}
There are three key properties of quantum mechanics that are used in the design of quantum computers: \emph{superposition}, \emph{interference} and \emph{entanglement}.
These properties allow them to represent and process information in ways that differ fundamentally from classical computers [\cite{pqc_migration_handbook}].
To help the reader grasp the essence of quantum properties, we quote Koen Groenland's illuminating introduction to superposition: \textit{`Think about any property that we can (classically) measure, such as the position of a particle or the value of a bit on a hard drive (0 or 1). 
In quantum mechanics, many different measurement outcomes can be somewhat ‘true’ at the same time: a particle can be in multiple positions at once, or a bit could be 0 and 1 simultaneously'} [\cite[ch.~1]{groenland}].
\paragraph{The quantum threat.}
Quantum computers leverage properties of quantum mechanics to solve certain problems far faster than classical computers.
While many potential applications are being researched, such as materials and optimisation [\cite{groenland}], their clearest application is breaking the mathematical assumptions behind currently deployed asymmetric cryptography [\cite{groenland,timelines}].
This is known as the \textit{quantum threat}, and quantum computers which are advanced enough to break encryption are pre-emptively named \textit{cryptographically relevant quantum computers} (CRQCs) [\cite{timelines}].
Such devices could run \emph{Shor’s algorithm} to solve the factoring and discrete-logarithm problems on which the asymmetric schemes rely \textit{efficiently}, enabling attackers to decrypt or forge data [\cite{groenland,pqc_migration_handbook}]. 
CRQCs are considered unlikely to emerge within ten years but highly likely within fifty [\cite{timelines}].
\paragraph{Quantum-safe cryptography.}
Quantum-Safe Cryptography (QSC) includes all cryptographic methods expected to withstand both classical and quantum attacks. 
This encompasses strengthened symmetric schemes and new cryptographic schemes such as Post-Quantum Cryptography (PQC) [\cite{pqc_migration_handbook}]. 
PQC is a major subset of QSC and focuses on new mathematical constructions, such as \emph{lattice-} or \emph{code-based} schemes specifically designed to resist quantum computers [\cite{pqc_migration_handbook}]. 
PQC runs on classical computers and does not rely on quantum properties [\cite{nist2035}].
Although PQC formally excludes existing symmetric algorithms, the term PQC is often used to refer to all cryptography expected to withstand quantum attacks [\cite{rfc}].

\paragraph{Quantum cryptography and QKD.}
In contrast to PQC, quantum cryptography encodes classical information into \textit{quantum states}, leveraging the quantum properties listed above [\cite{groenland}].
The most prominent example of quantum cryptography is Quantum Key Distribution (QKD) which can in theory provide a strong form of security even against attackers equipped with a quantum computer, unlimited time and unlimited computing power, meaning QKD would provide protection against the quantum threat [\cite{groenland}].
However, the high cost and technical complexity of QKD severely limit its use to specialised, high-security applications.
Despite these limitations, QKD remains an active area of research [\cite{verschoor_qkd_use_cases}].
\paragraph{NIST competition.}
In 2016, the US National Institute of Standards and Technology (NIST) initiated an international competition to develop and standardise post-quantum cryptographic schemes [\cite{nist2035}]. 
The first standards were released in 2024 and are now influencing global efforts to transition toward quantum-resistant encryption [\cite{pqc_migration_handbook}].
The White House subsequently mandated that all federal agencies must migrate to PQC by 2035, and NIST extended the competition with alternative candidates under consideration for future rounds [\cite{nist2035}].
\subsection{\textit{Boundaries and dynamics of innovation systems}}
As several scholars have noted, the authority and capability to drive large-scale sociotechnical transitions rarely rest with a single set of actors [\cite{Beck07022022, chesbrough2003open, geels2002technological}]. 
The complex challenges and `Catch-22' dynamics described in the quantum transition literature [\cite{KONG2024101884}] indicate that effective responses to the quantum threat emerge not from isolated interventions but through institutional co-evolution [\cite{geels2002technological}].
To capture this collective response to disruptive technological change, we adopt a systemic perspective on innovation, specifically through an innovation-system lens [\cite{Lundvall01022007,Wagner01061999}]. 
Given the central role of states in the adoption and regulation of cryptographic schemes, the national level provides an analytically useful scale for observing coordination processes.
The innovation system framework has been widely applied in the study of innovation and technological change, and was conceptualised by Lundvall as
\textit{`constituted by elements and relationships that interact in the production, diffusion and use of new and economically useful knowledge'}, and a National Innovation System (NIS) as encompassing \textit{`elements and relationships, either located or rooted inside the borders of a nation state'} [\cite[p.~86]{nis_def}].
Importantly, innovation-system theory recognises that the boundaries of a system are not fixed; rather, the inclusion or exclusion of actors is negotiated through ongoing social and institutional processes [\cite{Lundvall01022007,nelson1993nis,Wagner01061999}].
Modern perspectives such as \cite{van_der_Loos02012024} and \cite{Gifford28052021} apply innovation system perspectives to shape the development of governance in response to societal disruption.

The Netherlands provides a compelling case for our analysis of this disruption, as it is regarded as an advanced actor in the QSC migration, reflected in its participation in national initiatives such as the PCSI PQC benchmarking project\footnote{\scriptsize{\url{https://pcsi.nl/en/projects/pqc-benchmarking/}}}, its central role in the European Union’s response to the quantum threat through the EU PQC Workstream of the NIS Cooperation Group\footnote{\scriptsize{\url{https://digital-strategy.ec.europa.eu/en/library/coordinated-implementation-roadmap-transition-post-quantum-cryptography}}} and its contributions to transition literature [\cite{pqc_migration_handbook}].
Examining the National Innovation System of Quantum-Safe Cryptography in the Netherlands (\hyperlink{nisqscn}{NISQSCN}) thus provides an opportunity to identify barriers associated with this transition and to generate insights relevant to other national innovation systems undertaking similar quantum transitions. 
We thus define the \hyperlink{nisqscn}{NISQSCN} as our empirical object of analysis.

We adopt a structural approach, conceptualising the \hyperlink{nisqscn}{NISQSCN} as a network of interacting actors and institutions.
Focussing on mapping system composition, identifying government agencies, research institutes, private companies, standards bodies, and other relevant actors [\cite{KONG2024101884}], this approach clarifies the complex, relational nature of the innovation system [\cite{ANDERSSON2023100786}].
The socio-institutional dynamics of the \hyperlink{nisqscn}{NISQSCN} concern how these actors engage with regulations, norms, funding mechanisms, and trust frameworks to shape the governance and adoption of quantum-safe technologies [\cite{lundvall1992nis,nelson1993nis}]. 
Hence, our first research question is:  
\textit{Who are the key organisational actors involved in the development and deployment of QSC in the Netherlands?}
This framing positions the transition to QSC as a sociotechnical process embedded in networks of influence, responsibility, and institutional coordination, shaped not only by system configuration but also by conditions that enable or constrain collective action.
\subsection{\textit{Responsibilities, trust and organisational dynamics}}\label{trust}
We extend the structural framing established above to a functional one as a framework for analysing how diverse actors manage the extensive distributed coordination and institutional change required in this transition.
We do not adopt a full functional approach, but a \textit{functional-inspired} approach, focusing on identifying the actual and intended roles and responsibilities in order to examine how actors influence system development and identify the barriers that inhibit them [\cite{lundvall1992nis,nelson1993nis,HEKKERT2007413}].
By revealing where gaps remain, our work uncovers functions within the \hyperlink{nisqscn}{NISQSCN}, offering a direct route to targeted and actionable policy insights.
Previous research has already identified transition policy priorities, for instance \cite{KONG2024101884} present specific recommendations for governmental action.
Our analysis extends the scope beyond government to examine how roles, responsibilities, and expertise are distributed across the full range of actors in the \hyperlink{nisqscn}{NISQSCN}.
Hence, our second research question is: \emph{What roles and responsibilities do these actors assume in supporting the transition to QSC?}

Applying a functional-inspired approach deepens our systemic understanding of the \hyperlink{nisqscn}{NISQSCN} by clarifying how expertise, accountability, and influence are allocated across the innovation system, essential insights for steering coordinated transition. 
Yet coordination across these roles depends on more than institutional design: it also relies on maintaining trust among actors engaged in standardisation and governance.
The reliability of information surrounding quantum technologies is not guaranteed, prompting community-led efforts to filter misinformation, such as expert-curated platforms that flag unreliable claims\footnote{\scriptsize{\url{https://www.wired.com/story/revolt-scientists-say-theyre-sick-of-quantum-computings-hype/}}}.
Trust remains central because the integrity of cryptographic infrastructures depends not only on technical robustness but also on credible governance. 
Past failures such as the Dual EC random number generator controversy illustrate how opacity in decision-making can erode confidence in international standards\footnote{\scriptsize{\url{https://www.niskanencenter.org/the-nsa-and-nist-a-toxic-relationship/}}}, which demonstrates that trust is not ancillary but constitutive of governance decisions.
By contrast, the Internet Engineering Task Force (IETF), where highly transparent standardisation processes ensure broad participation and consensus-driven decisions\footnote{\scriptsize{\url{https://www.ietf.org/process/}}}, exemplifies how trust can be built and maintained.
Through these examples, we underscore how inclusivity and transparency reinforce the legitimacy of technical standards, and note that trust underpins the system function of \textit{Legitimation} [\cite{HEKKERT2007413}].
Beyond shaping current governance arrangements, trust also conditions how actors imagine and coordinate toward desired technological futures, discussed in the following section.
\subsection{\textit{Sociotechnical imaginaries, futures and transition pathways}}\label{imaginaries}
\begin{figure}
    \centering
    \includegraphics[width=0.4\linewidth]{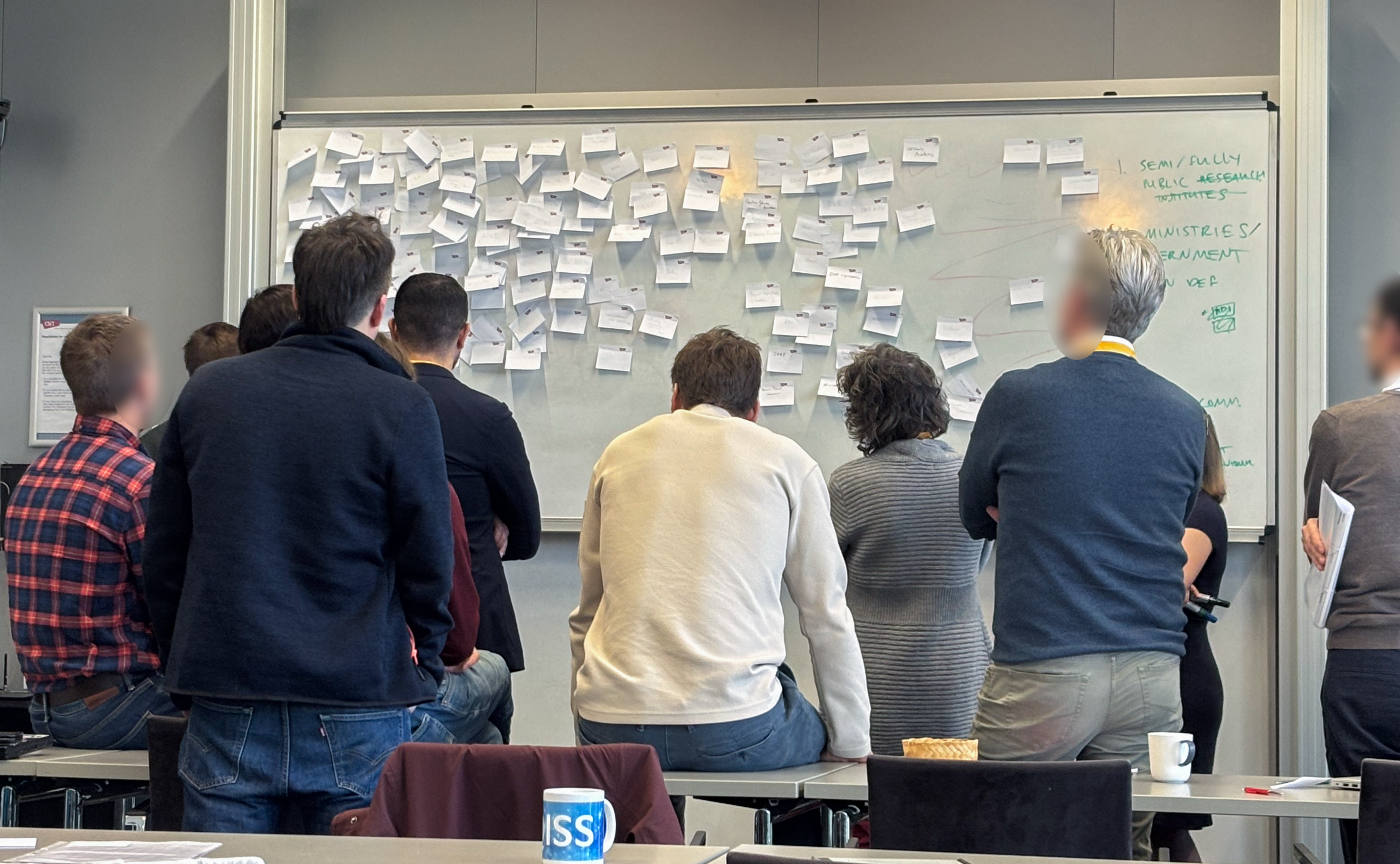}
    \caption{The first exercises of the workshop took place in plenary.}
    \label{fig:grouping}
\end{figure}
To navigate the networked dependencies which characterise the quantum transition, we build on the functional and trust-based dynamics by drawing on the concept of \emph{sociotechnical imaginaries} to articulate collective expectations of a quantum-safe future [\cite{jasanoff2009dreamscapes, gerhold2021imaginaries}].
\footnote{We adopt imaginaries rather than scenarios because our aim is to analyse normative visions shaping governance, not to forecast technical developments.}
Jasanoff and Kim define sociotechnical imaginaries as \textit{`collectively held, institutionally stabilized, and publicly performed visions of desirable futures, animated by shared understandings of forms of social life and social order attainable through, and supportive of, advances in science and technology' [\cite[p.~4]{jasanoff2009dreamscapes}}].
In essence, sociotechnical imaginaries describe how societies envision their futures in relation to technological change and how these visions in turn shape trajectories of innovation and governance.

Sociotechnical imaginaries structure expectations among system actors, influencing coordination by aligning, or indeed misaligning, perceived responsibilities and timelines.
They actively structure present-day decisions, shaping institutional priorities, regulatory design, and coordination practices by embedding normative assumptions about what a secure and responsible quantum future should look like.
Such imaginaries are visible, implicitly or explicitly, in contemporary policy discourses on post-quantum migration [\cite{Csenkey01092024,KONG2024101884}]. 
Futures-oriented perspectives, such as Beckman’s work on systems leadership in 2053, further underscore the importance of anticipatory governance and long-term stewardship of innovation systems [\cite{beckman2024designing}].
At the broadest level, the prevailing sociotechnical imaginary in this domain envisions that migration to QSC will be completed before the advent of large-scale quantum computing.
Given this collectively held goal and the dependence of transition success on distributed coordination, our third research question is: \emph{How can the (Dutch) transition to a quantum-safe innovation system be managed proactively?}
We build on these conceptual foundations to consider how sociotechnical dynamics translate into governance arrangements and policy pathways for managing the quantum-safe transition. 

Transition governance requires navigating complex dynamics, including power asymmetries, institutional structures, and networked interdependencies, factors which influence how governance arrangements evolve over time as boundaries and responsibilities across sectors are continuously negotiated [\cite{geels2002technological, Stirling31122024}]. 
Hence, the transition must be understood in the broader context of national policy transformations, where trust and legitimacy are as critical as technical standards.
Finally, this transition must also be situated beyond the national setting. 
While the Dutch case provides valuable insight, it is embedded within a European landscape of parallel transitions, where multilevel coordination and shared trust remain enduring challenges. 
This European context is itself embedded in a global environment where geopolitical dynamics further complicate the governance of cryptographic standards and technological futures. 
Recognising these interconnected contexts grounds the subsequent analysis and provides the basis for identifying effective governance and transition pathways later in the paper.
\section{Methodology}
The workshop took place at Centrum Wiskunde \& Informatica (CWI) 
in Amsterdam Science Park on 29 January 2025. 
Fifteen experts from industry, academia and the Dutch government participated in a four-hour session to scope the \hyperlink{nisqscn}{NISQSCN} and identify barriers to its development.
Participants reported that the event provided useful opportunities for discussion, insight, and networking.
This workshop employed a qualitative, iterative focus group design, in which participants co-constructed the innovation system, enabling a deep discussion; we reflect on this in \cref{sec:summary}.
Limitations include the potential bias introduced by participants, and the limited generalisability of this approach due to the deeply context-specific findings which may be less applicable to other populations.
To mitigate these limitations, future research could use a mixed-methods approach, combining qualitative insights with quantitative data to improve generalisability, and expanding the participant pool to include a more diverse range of stakeholders could help reduce bias and provide a broader understanding of the innovation system.
We describe the study design, including participant selection, data-collection procedures, and the analytical approach used after the event.
\begin{table}[ht]
\footnotesize
\centering
\begin{tabularx}{\textwidth}{l p{0.35\textwidth} p{0.2\textwidth}}
\toprule
\textbf{Organisation} & \textbf{Primary Role} & \textbf{Label} \\
\midrule
Dutch Banking Association & Cybersecurity Policy Advisor & Supervisor-1 \\
Dutch Inspectorate for Digital Infrastructure & Coordinating/Specialist Inspector & Supervisor-2 \\
Ministry of Interior Affairs & Digital Resilience Program Manager & Government-1 \\
Ministry of Economic Affairs & Policy Officer & Government-2 \\
Ministry of Defence & Senior Innovation Manager & Government-3 \\
Government Quantum Threat Working Group & Program Manager &  Government-4 \\
Research Organisation & Cryptology Researcher & Research-1 \\
Research Organisation & Director & Research-2 \\
Research Organisation & Cryptology Researcher & Research-3 \\
IT Service Management Company & Research Engineer & Research-4 \\
QSC Migration Service Provider & Security Expert \& Advisor & Industry-1 \\
Quantum Technology Ecosystem & Ecosystem Development Coordinator & Promoter-1 \\
Consulting Firm & Cybersecurity Consultant & Advisor-1 \\
Dutch Cybersecurity Agency & Cybersecurity Advisor & Advisor-2 \\
Dutch National Bank & Operational \& IT Risk Supervisor & Regulator-1 \\
\bottomrule
\end{tabularx}
\caption{Affiliation and primary role of workshop participants.}
\label{tab:workshop_participants}
\end{table}
\subsection{\textit{Study design}}
\begin{figure}[htbp]
    \centering
        \begin{tikzpicture}[scale=0.7]
        \begin{axis}[
            view={20}{30},
            xlabel={business},
            ylabel={research},
            zlabel={policy},
            grid=both,
            major grid style={line width=0.8pt,draw=gray!60},
            minor grid style={dotted,draw=gray!30},
            minor tick num=4
        ]
        \addplot3[
            only marks,
            mark=*,
            blue
        ]
        table[
            x=business,
            y=research,
            z=policy,
            col sep=comma
        ] {data.csv};
        \end{axis}
    \end{tikzpicture}
    \qquad
    \begin{tikzpicture}[scale=0.7]
        \begin{axis}[
            view={80}{30},
            xlabel={business},
            ylabel={research},
            zlabel={policy},
            grid=both,
            major grid style={line width=0.8pt,draw=gray!60},
            minor grid style={dotted,draw=gray!30},
            minor tick num=4,
        ]
        \addplot3[
            only marks,
            mark=*,
            blue
        ]
        table[
            x=business,
            y=research,
            z=policy,
            col sep=comma
        ] {data.csv};
        \end{axis}
    \end{tikzpicture}
    \caption{Perceived expertise of workshop participants, shown from two angles. Interactive visualisation available at \url{https://quantumimpact.shinyapps.io/expertise/}.}
    \label{fig:visualisation}
\end{figure}
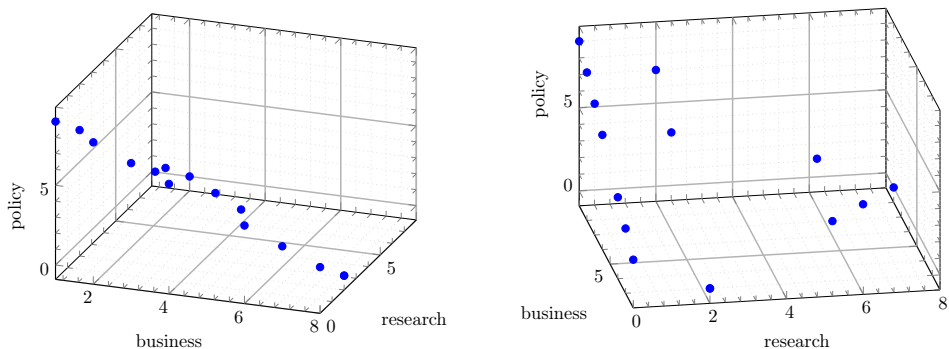

As the workshop aimed to scope the \hyperlink{nisqscn}{NISQSCN}, participant selection was designed to capture diverse perspectives across the triple helix of industry, academia and government. 
With no comprehensive list of quantum threat experts available, participants were identified iteratively in consultation with peer academics. 
Industry expertise is concentrated among a small set of cybersecurity professionals active in public forums and at industry events, while academic expertise is broader and traceable through research outputs which relate to the quantum threat. 
Government participants were drawn from agencies involved in an interdepartmental working group on the quantum threat.
An initial list of twelve candidates was compiled to balance expertise and institutional roles.
Each candidate’s experience was assessed across three domains (business, research and policy), and visualised to ensure a balanced distribution (\cref{fig:visualisation}). 
Peer feedback was used to refine the list and fill gaps, including adding a civil society representative.
When invitations were declined, substitutes with comparable backgrounds were recruited.
The final workshop included fifteen participants (\cref{tab:workshop_participants}), reached through purposive and convenience-based sampling within professional networks, with external participants approached via mutual introductions.
During the workshop, participants were assigned to breakout groups to distribute expertise evenly across thematic areas (\cref{tab:breakout_allocations}). 
For analytical clarity, participants are assigned labels reflecting their primary role.

\subsection{\textit{Data collection}}
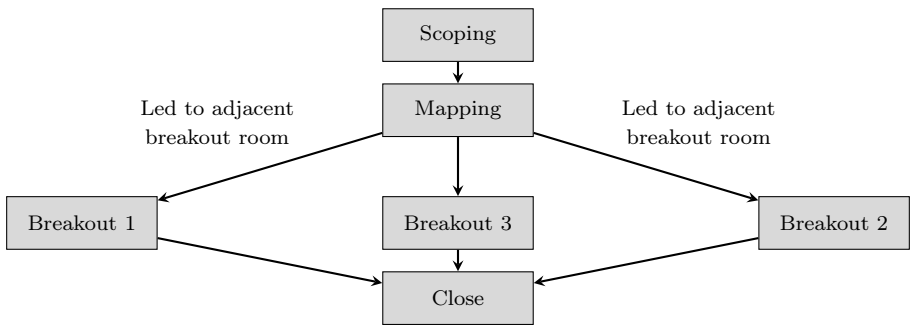
\begin{figure}[ht]
    \centering 
        \begin{tikzpicture}[node distance=1cm]
            \node (scoping) [process] {\footnotesize Scoping};
            \node (mapping) [process, below of=scoping] {\footnotesize Mapping};
            \node (b3) [process, below of=mapping, yshift = -0.5cm] {\footnotesize Breakout 3};
            \node (b1) [process, left of=b3, xshift=-4cm] {\footnotesize Breakout 1};
            \node (b2) [process, right of=b3, xshift=4cm] {\footnotesize Breakout 2};
            \node (close) [process, below of=b3] {\footnotesize Close};
            \node (annotation1) [left of=mapping, xshift=-2.2cm] {\footnotesize Led to adjacent};
            \node (annotation2) [left of=mapping, xshift=-2.2cm, yshift=-0.35cm] {\footnotesize breakout room};     
            \node (annotation3) [right of=mapping, xshift=2.2cm,] {\footnotesize Led to adjacent};
            \node (annotation4) [right of=mapping, xshift=2.2cm, yshift=-0.35cm] {\footnotesize breakout room};
            \draw [arrow] (scoping) -- (mapping);
            \draw [arrow] (mapping) -- (b3);
            \draw [arrow] (mapping) -- node[anchor=east] {} (b1);
            \draw [arrow] (mapping) -- node[anchor=east] {} (b2);
            \draw [arrow] (b3) -- (close);
            \draw [arrow] (b1) -- node[anchor=east] {} (close);
            \draw [arrow] (b2) -- node[anchor=east] {} (close);
        \end{tikzpicture}
        \caption{Experimental study design flowchart.}
    \label{fig:study_design_flowchart}
\end{figure}

The half-day workshop at CWI in Amsterdam used one plenary room and two breakout rooms to delineate the boundaries and actors of the \hyperlink{nisqscn}{NISQSCN} collaboratively. 
As shown in \cref{fig:study_design_flowchart}, the programme comprised four activities: Scoping, Mapping, Breakouts and Closing. 
After an informal pre-session interaction, the plenary Scoping session introduced the study objectives, aligned terminology (PQC, QKD, QSC; see \cref{sec:key_terms}), and presented a definition of cryptoagility. 
Consensus on the scope and definition of the \hyperlink{nisqscn}{NISQSCN} was established before moving to Mapping, where participants individually listed organisations they considered important in the \hyperlink{nisqscn}{NISQSCN}.
These were collectively grouped and labelled through discussion, producing an agreed set of categories that represented the perceived structure of the national innovation system.
Outputs were documented photographically.
Participants were then divided into three breakout groups, each addressing a subset of the identified organisations. 
Guided by a leader and a facilitator, groups discussed the \textit{intended} and \textit{actual} responsibilities of each organisation. 
Breakout discussions were audio recorded and all visual materials were archived for analysis, with participants informed in advance about the recording procedures.
The workshop concluded with a Closing plenary, during which key insights were summarised, feedback was provided by participants and next steps were discussed.
A workshop report was subsequently shared with participants.
We highlight that throughout this work respondents are referred to in their professional capacities, not in terms of their personal opinion, regardless of how informed by their expertise this may be.
\subsection{\textit{Thematic coding}}
In terms of analysis, we applied a thematic coding to the transcripts and notes from the workshop discussions in order to perform \textit{thematic analysis} [\cite{Braun01012014}]. 
The initial set of themes (presented in \cref{tab:codeschema}) was developed collaboratively during the workshop itself, as part of a scoping discussion with participants, whom we treated as peer researchers. 
These themes reflect the roles and responsibilities identified by stakeholders across the \hyperlink{nisqscn}{NISQSCN} and served as a shared conceptual starting point for the analysis.
Following the coding process, we compared our independently generated code sets and calculated an intercoder reliability score.
In qualitative research, intercoder reliability values of 0.80 or higher are typically considered acceptable, indicating substantial agreement between coders [\cite{intercoder_reliability}].
A percent-agreement score at this level suggests that the researchers independently identified a highly similar set of themes and subthemes; our score was 0.88, which falls well within the acceptable range.
Where discrepancies arose, we discussed and refined our interpretations until consensus was reached. 
Our analysis continued until theoretical saturation was achieved; that is, when no new themes were emerging from the data and existing categories were sufficiently rich and well-defined. 
While our initial themes shaped the coding process, we remained open to emergent insights. 
For example, the theme of \textit{role tensions and misalignment} was not explicitly raised in the Scoping discussion but became analytically significant through patterns of ambiguity, contradiction, and overlap in how responsibilities were described and negotiated. 
This flexible approach allowed us to capture both the formal and situated dynamics within the \hyperlink{nisqscn}{NISQSCN}.
\begin{table}[ht]
\footnotesize
\centering
\begin{tabularx}{\textwidth}{X X X X}
\toprule
\textbf{Role in Breakout} &\textbf{Breakout 1} & \textbf{Breakout 2} & \textbf{Breakout 3} \\
\midrule
Lead &Lead-1 & Lead-2 & Lead-3 \\
Facilitator &Facilitator-1 & Facilitator-2 & Facilitator-3 \\
Participant &Government-2 & Government-1 & Government-4 \\
Participant &Research-4 & Research-2 & Research-1 \\
Participant &Supervisor-2 & Advisor-2 & Government-3 \\
Participant &Industry-1 & Research-3 & Advisor-1 \\
Participant &Promoter-1 & Regulator-1 & Supervisor-1 \\
\bottomrule
\end{tabularx}
\caption{Allocation of participants to breakout groups.}
\label{tab:breakout_allocations}
\end{table}
\subsection{\textit{Researcher reflexivity}}
We now apply reflexivity to our research by critically examining potential sources of bias and reflecting on how our positionalities and prior experiences informed and influenced the analytical process.
The first author, affiliated with the University of Amsterdam (UvA) and QuSoft, had existing professional relationships with several participants through national and European quantum policy work, providing contextual insight but also potential interpretative bias.
The second author, also affiliated with UvA, contributed critical research expertise and advised on the qualitative methodology, drawing on a background in sociotechnical systems and innovation studies. 
Together, these positionalities deepened contextual insight while requiring reflexivity to account for how prior relationships may influence our understanding of the data.
The workshop was designed as a respondent-led exercise, with participants contributing to the framing of innovation-system boundaries and shared terminology. 
We conducted reflexive thematic analysis, coding independently and discussing emergent patterns iteratively. 
Our interpretations are shaped by both the data and our embeddedness in the field, which we acknowledge as a beneficial resource, enabling informed analysis, as well as a limitation, introducing partiality.
\begin{table}[htbp]
\centering
\rotatebox{90}{%
\begin{minipage}{\textheight}
\footnotesize
\renewcommand{\arraystretch}{1.3}
\caption{Thematic coding schema for actor roles in QSC innovation system.}
\label{tab:codeschema}
\begin{tabular}{>{\raggedright\arraybackslash}p{2.8cm} >{\raggedright\arraybackslash}p{3cm} >{\raggedright\arraybackslash}p{6cm} >{\raggedright\arraybackslash}p{6.6cm}}
\toprule
\textbf{Code} & \textbf{Theme} & \textbf{Definition} & \textbf{Example (quotation)} \\
\midrule
Tech\_Dev & Technical Development & Involvement in algorithm design, cryptographic tools, security audits, etc. & `[A role of research and education is to] fill the gap, invent or research new cryptographic algorithms or protocols.' \\
Implementation & Implementation \& Integration & Activities around adopting, integrating, or migrating to quantum-safe systems. & `[A role of End Users is to] adopt or implement solutions.' \\
Regulation\_Policy & Regulation \& Policy & Involvement in setting rules, standards, legal frameworks, or compliance processes. & `If [regulators] can put a dot on the horizon, the branches can... calculate when they have to start their transition.' \\
Funding & Funding \& Incentivisation & Provision or allocation of financial resources, funding programmes, or incentives. & `[Financiers] [put] money on the table to create solutions.' \\
Standards\_Gov & Standards \& Governance & Participation in formal standardisation and governance bodies. & `[Organisations] might decide \textit{I don't care about the confidentiality of my users} but [supervisors] intervene and they are forced to care.' \\
Awareness & Awareness-Raising \& Communication & Efforts to educate, inform, or build awareness internally or externally. & `Creating awareness... that's what [promoters] do, and they do that by knowledge sharing.' \\
Coordination & Coordination \& Convening & Facilitation of stakeholder collaboration, alignment, and strategic coordination. & `Recent PQ [consortia] have done a really good service bringing people together.' \\
Expertise & Expertise \& Interpretation & Expressions of expertise, authority, or interpretation of technical relevance. & `[Regulators] will set a goal, and they don't know how the goal will be reached, and [supervisors] write the regulations... the way that you can actually fulfil your goal.' \\
Gaps & Responsibility Gaps or Ambiguity & Mentions of unclear or unclaimed responsibilities, delays, or organisational blind spots. & `Why would [end users] now make all this effort... and then different standards come out and [they] might not be compliant?'  \\
Tension & Role Tensions or Misalignment & Conflicts, overlaps, or disagreements over who should be responsible for what. & 
`The raison d'être [of standardisation bodies] [is to] make the standards themselves, so it's a bit weird to co-ordinate with each other.'
 \\
\bottomrule
\end{tabular}
\end{minipage}
}
\end{table}
\section{Results}
Using a reflexive thematic analysis that combined workshop-generated categories with inductively developed themes [\cite{Braun01012014}], we present findings aligned with our three research questions. 
\cref{mdis} examines how participants defined system boundaries and identified key organisational actors (\ref{RQ1}).
\cref{arr} analyses perceived formal and interpretative roles and responsibilities, as well as where tensions and gaps emerged (\ref{RQ2}). 
\cref{mt} builds on our analysis of the interpretative roles to explore participants’ imaginaries of the future system, focussing on how questions of readiness, standards, and trust were framed as central to managing the transition (\ref{RQ3}).
\subsection{\textit{Mapping the Dutch innovation system of QSC}}\label{mdis}
\subsubsection{Scoping and mapping exercises}
The first activity involved aligning on key terminology, a necessary step to establish shared understanding across diverse stakeholders and to strengthen the internal validity of subsequent discussions.
We reiterate that the meanings of some key terms shift across contexts, and misalignment can occur even among experts.
Participants first aligned on the notions of traditional cryptography, PQC, QKD, QSC and hybrid cryptography.
Definitions of the first three terms were unanimously accepted and are listed in \cref{sec:key_terms}. 
Hybrid cryptography was accepted as the combination of traditional cryptography with PQC.
Next, the group reviewed a proposed definition of cryptoagility, slightly adapted from the 2024 PQC Migration Handbook [\cite{pqc_migration_handbook}],
\emph{Cryptographic agility enables organisations to quickly modify or replace deployed cryptographic primitives without significantly disrupting organisational processes.
Software, hardware, policies and departments can be cryptoagile.}
This definition was accepted without any suggested amendments, signalling unanimous consensus among the participants.  

With key terms established, a definition of the innovation system could be proposed. 
Initially, the system was framed in terms of \textit{active roles} in the adoption of PQC, QKD, hybrid cryptography, and cryptoagility. 
The final accepted definition was revised to focus on accountability and responsibility:
\emph{The innovation system of QSC in the Netherlands consists of organisations responsible or accountable (e.g., developing, testing, implementing, regulating, raising awareness, financing) for the migration to PQC, QKD, or hybrid cryptography.}  
Notable revisions from the original proposed definition include (i) emphasis placed on accountability rather than activity; (ii) the inclusion of financing and awareness-raising as core functions; (iii) the exclusion of cryptoagility, a divisive decision which we examine below. 
We note that (i) indicates a normative framing, and that the activities judged by the group to be important enough to include explicitly in the ecosystem definition centre on steering (including financing) and soft conferral of responsibility (awareness-raising), with advanced technical aspects (cryptoagility) of little enough importance that they should be removed.
We elaborate on the key boundary decisions taken and reflect on their implications.
\begin{figure}[htbp]
    \centering 
        \begin{tikzpicture}[node distance=1.3cm]
            \node (promoters) [technology] {\scriptsize{Promoters}};
            \node (financiers) [technology, below of=promoters, yshift=-0.3cm] {\scriptsize{Financiers}};
            \node (research) [technology, right of=promoters, xshift=2cm] {\begin{singlespace}
                \scriptsize{Research \& education}
            \end{singlespace}};
            \node (service) [technology, right of=research, xshift=2cm] {
            \begin{singlespace}
                \scriptsize{Migration service providers}
            \end{singlespace}};
            \node (manufacturers) [technology, above of=service, yshift=0.7cm] {\begin{singlespace}
                \scriptsize{Manufacturers of cryptography components}
            \end{singlespace}};
            \node (networks) [technology, below of=service, yshift=-0.5cm] {\begin{singlespace}
                \scriptsize{Network operators}
            \end{singlespace}};
            \node (consulting) [technology, right of=service, xshift=2cm] {\begin{singlespace}
                \scriptsize{Consulting \& advisory}
            \end{singlespace}};
            \node (standards) [standards, below of=research, yshift=-2.4cm] {\begin{singlespace}
                \scriptsize{Standardisation Bodies}
            \end{singlespace}};
            \node (regulators) [standards, right of=standards, xshift=2cm, yshift=0.5cm]{\scriptsize{Regulators}};
            \node (supervisors) [standards, below of=regulators] {\scriptsize{Supervisors}};
            \node (branch) [standards, right of=supervisors, xshift=2cm] {\begin{singlespace}
                \scriptsize{Branch organisations}
            \end{singlespace}};
            \node (users) [users, right of=networks, xshift=4cm, yshift=-1cm] {\scriptsize{End Users}};
            \node (test0) [right of=consulting, xshift=-0.5cm, yshift=1.8cm] {};
            \node (test1) [above of=test0, yshift=-0.8cm] {};
            \node (test2) [right of=test0] {};
            \node (test3) [right of=test1] {};
            \draw [dasharrow] (regulators) -- (users);
            \draw [dasharrow] (branch) -- (users);
            \draw [doubledasharrow] (supervisors) -- (regulators);
            \draw [doubledasharrow] (standards) -- (regulators);
            \draw [dasharrow] (supervisors) -- (users);
            \draw [arrow] (promoters) -- (research);
            \draw [arrow] (financiers) -- (research);
            \draw [arrow] (research) -- (networks);
            \draw [arrow] (research) -- (manufacturers);
            \draw [arrow] (research) -- (service);
            \draw [dasharrow] (manufacturers) -- (service);
            \draw [doublearrow] (service) -- (consulting);
            \draw [dasharrow] (consulting) -- (users);
            \draw [dasharrow] (service) -- (users);
            \draw [arrow] (research) -- (service);
            \draw [dasharrow] (networks) -- (users);
            \draw [dasharrow] (test0) -- (test2);
            \draw [arrow] (test1) -- (test3);
            \draw [dasharrow] (research) edge [bend left=22] (consulting);
            \draw [arrow] (financiers) -- (service);
            \node (annotation1) [right of=regulators, xshift=1.3cm, yshift=0.4cm] {\tiny{\hyperlink{R5}{R5}, \hyperlink{R6}{R6}, \hyperlink{R18}{R18}, \hyperlink{R19}{R19}}};
            \node (annotation2) [right of=branch, xshift=0.1cm, yshift=0.9cm] {\tiny{\hyperlink{R1}{R1}}};
            \node (annotation3) [right of=standards, xshift=-0.1cm, yshift=0.6cm] {\tiny{\hyperlink{R20}{R20}, \hyperlink{R24}{R24}, \hyperlink{R25}{R25}}};
            \node (annotation4) [right of=promoters, xshift=0.3cm, yshift=0.15cm] {\tiny{\hyperlink{R17}{R17}}};
            \node (annotation5) [above of=financiers, xshift=0.9cm, yshift=-0.5cm] {\tiny{\hyperlink{R10}{R10}, \hyperlink{R11}{R11}}};
            \node (annotation6) [right of=research, xshift=0.3cm, yshift=0.2cm] {\tiny{\hyperlink{R23}{R23}}};
            \node (annotation7) [right of=networks, xshift=1.1cm, yshift=-0.3cm] {\tiny{\hyperlink{R8}{R8}, \hyperlink{R16}{R16}}};
            \node (annotation8) [right of=research, xshift=-0.1cm, yshift=1cm] {\tiny{\hyperlink{R23}{R23}}};
            \node (annotation9) [right of=service, xshift=0.2cm, yshift=0.2cm] {\tiny{\hyperlink{R4}{R4}}};
            \node (annotation10) [right of=consulting, xshift=0.5cm, yshift=-1.6cm] {\tiny{\hyperlink{R2}{R2}, \hyperlink{R3}{R3}, \hyperlink{R9}{R9}}};
            \node (annotation11) [below of=consulting, xshift=-0.3cm, yshift=0.2cm] {\tiny{\hyperlink{R7}{R7}, \hyperlink{R14}{R14}, \hyperlink{R15}{R15}}};
            \node (annotation12) [right of=supervisors, xshift=1.0cm, yshift=1.0cm] {\tiny{\hyperlink{R27}{R27}, \hyperlink{R28}{R28}}};
            \node (annotation13) [left of=networks, xshift=-0.5cm, yshift=0.7cm] {\tiny{\hyperlink{R23}{R23}}};
            \node (annotation14) [below of=manufacturers, xshift=0.3cm, yshift=0.4cm] {\tiny{\hyperlink{R13}{R13}}};
            \node (annotation15) [below of=regulators, xshift=-0.9cm, yshift=0.5cm] {\tiny{\hyperlink{R21}{R21}, \hyperlink{R26}{R26}, \hyperlink{R29}{R29}}};
            \node (annotation16) [right of=financiers, xshift=1.2cm, yshift=0.4cm] {\tiny{\hyperlink{R12}{R12}}};
            \node (annotation17) [left of=consulting, xshift=-0.1cm, yshift=0.8cm] {\tiny{\hyperlink{R22}{R22}}};
            \node (annotation18) [right of=test2, xshift=-0.8cm] {\tiny{Intended}};
            \node (annotation19) [right of=test3, xshift=-0.85cm] {\tiny{Fulfilled}};
            \end{tikzpicture}
    \caption{Visualisation of the Dutch QSC innovation system.
    The figure maps the responsibilities identified in \cref{tab:roles_responsibilities_from_transcripts} onto the relevant actors, illustrating two parallel flows, technology development (dark grey) and standardisation (light grey). 
    Solid lines denote QSC transition responsibilities that are currently well-functioning, while dashed lines indicate responsibilities that are intended but not yet fulfilled.}
    \label{fig:validated_actors}
\end{figure}
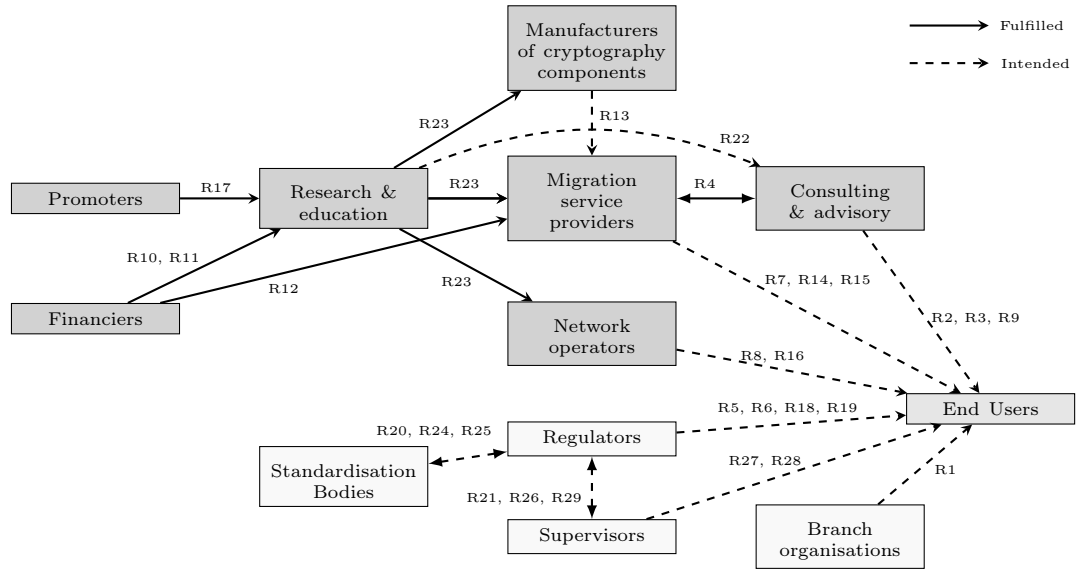

\subsubsection{Boundary decisions}
During the plenary Scoping discussion, one participant suggested to frame the discussion around the shared societal goal of resilience against quantum-enabled attacks, rather than to consider the \hyperlink{nisqscn}{NISQSCN} explicitly. 
While participants acknowledged this broader framing, it was deemed too expansive for the workshop's focus on QSC adoption. 
This debate nonetheless represents a normative perspective, as it is grounded in the societal value of maintaining cybersecurity, strengthening our notion that the \hyperlink{nisqscn}{NISQSCN} is normative.

As mentioned above, although cryptoagility features prominently in migration handbooks and policy guidance [\cite{pqc_migration_handbook}], and two participants considered it the most important aspect of transition, most participants emphasised during the Scoping discussion that it was not a relevant organising principle for the \hyperlink{nisqscn}{NISQSCN}.
This decision reflects a critical assessment of its practical value: several participants noted that cryptoagility has often led to weak implementations of otherwise strong protocols, while many organisations remain far from achieving even basic cryptoagile capabilities.
More broadly, the rejection of cryptoagility signals a pragmatic orientation within the innovation system, where technical ideals are evaluated against the realities of infrastructural ossification\footnote{\scriptsize{\url{https://www.ietf.org/slides/slides-semiws-ossification-a-result-of-not-even-trying-00.pdf}}}.

Participants also debated the relevance of the American Federal Government, particularly NIST, during the Scoping discussion. 
While NIST's PQC standardisation process was recognised internationally, its influence was perceived inconsistently at the national level.
Six participants described the American Federal Government or NIST as highly influential, shaping algorithms and migration trajectories, while others argued that it had little direct bearing on the Dutch context and should therefore not be treated as a central actor. 
These divergent views illustrate how actors in the \hyperlink{nisqscn}{NISQSCN} draw boundaries of relevance differently, reflecting both technical dependencies on global standards and a desire to foreground domestic institutional arrangements.

We now summarise the extended discussion received by a few actors.
First, participants identified a subset of End Users as `pioneers': organisations undertaking early migration efforts partly for prestige.
As one academic noted, successful transitions sometimes function as `a PR move more than an actual experiment', suggesting that pioneering was interpreted as a symbolic rather than strictly technical role.
Second, an actor group initially labelled \textit{Lobbyists} was renamed \textit{Promoters} during Mapping.
This actor group encompasses organisations which aim to prompt transition activities in other actors.
One participant summarised that the label had been changed to sound `more positive', although the motivation for this was not explicitly stated.
We conjecture that the positive re-framing was motivated not only by a desire to avoid criticising participants belonging to this group, but also by an effort to depoliticise the terminology and foster a more neutral or collaborative framing of the role.
Finally, we clarify that Dutch \textit{branch} organisations are representational bodies that unite organisations within specific sectors, e.g. the Dutch Banking Association (NVB).
These discussions imply that the innovation system is shaped not only by the technology but also by institutional and organisational dynamics.

Finally, we describe how the fifteen proposed system actors were validated.
In a short break between the Mapping and Breakout activities, the breakout leads reviewed the fifteen proposed actors and identified that three groups did not have distinct roles: some groups (such as the general public, big tech, and politicians) were incorporated into broader categories (e.g. End Users, Regulators).
During the subsequent Breakout discussions, participants validated the remaining twelve groups, which were accepted and are listed in \cref{fig:validated_actors}.
An industry expert identified a subcategory of End Users, described as Special End Users, characterised by having enough buying power that they have an influence on the innovation system.
Participants in Breakout 1 debated whether Network Operators had a unique role in the \hyperlink{nisqscn}{NISQSCN}, and concluded that they do have a unique role with respect to QKD, but they are simply End Users with regards to all other QSC.
This iterative validation process reflects that the experts did not perceive the boundary and composition of this innovation system as absolute, but as negotiated.
\begin{figure}
    \centering
    \includegraphics[width=0.4\linewidth]{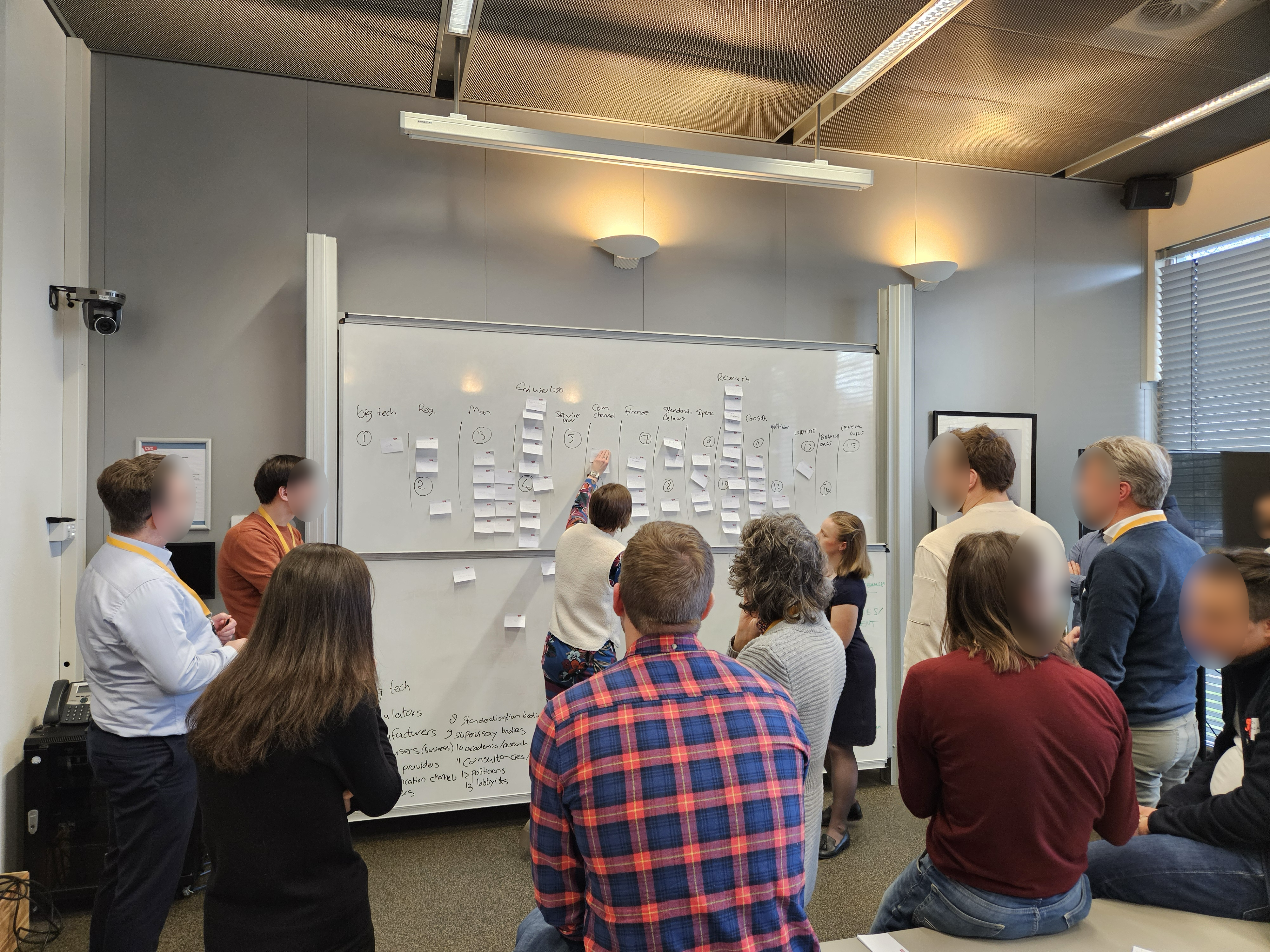}
    \caption{The research team repositioned system actors guided by participants' instructions.}
    \label{fig:editing_categories}
\end{figure}
\begin{table}[!h]
\centering
\rotatebox{90}{%
\begin{minipage}{\textheight}
\footnotesize
\renewcommand{\arraystretch}{1.3}
\caption{Roles and responsibilities of NISQSCN actors, both formal (\ref{RQ2}) and interpretative (\ref{RQ3}).}
\label{tab:roles_responsibilities_from_transcripts}
\begin{tabular}{>{\raggedright\arraybackslash}p{2cm} >{\raggedright\arraybackslash}p{9cm} >{\raggedright\arraybackslash}p{9cm}}
\toprule
 \textbf{Actor} & \textbf{Formal Roles} & \textbf{Interpretative Roles} \\
\midrule
Branch & 
Define best practice. &
Promote and incentivise best practice (\hypertarget{R1}{R1}). \\
\hline
Consulting \& Advisory & 
Implement solutions (\hypertarget{R2}{R2}); advise on quantum threat, migration, and incentives; advise on cryptography (\hypertarget{R3}{R3}); make partnerships with tools and migration service providers (\hypertarget{R4}{R4}). &
Win business; maintain compatibility with adjacent organisations.\\
\hline
End Users & 
Comply to existing standards/regulations (\hypertarget{R5}{R5}); 
incorporate quantum threat in risk assessment;
understand own needs;
express demands to vendors;
create migration strategy; 
create internal cryptography policy;
analyse urgency of use cases; 
estimate transition cost; 
allocate budget;
upskill employees;
adopt new standards/regulations (\hypertarget{R6}{R6}); 
develop internal policy;
identify migration barriers; 
adopt/buy tooling/services (\hypertarget{R7}{R7}, \hypertarget{R8}{R8}, \hypertarget{R9}{R9}); 
make cryptography inventory; 
implement; 
experiment/test; 
maintain cryptography and cryptoagility. 
& 
Recognise the quantum threat;
start migration in broad sense; 
create awareness internally; 
prevent internal silos; 
be willing to finance; 
contest managerial migration mandates;
express future demands to vendors. \\
\hline
Financiers & 
Finance fundamental research (\hypertarget{R10}{R10}); finance applied research (\hypertarget{R11}{R11}); finance startups (\hypertarget{R12}{R12}). & 
Promote innovation; steer funding towards important and non-profitable activity; distribute funding across system.\\
\hline
Manufacturers & 
Incorporate PQC in products/services (\hypertarget{R13}{R13}). & 
- \\
\hline
Migration Service Providers & Create and maintain scalable tooling/services; offer standard migration services (\hypertarget{R14}{R14}); offer bespoke migration services (\hypertarget{R15}{R15}).
& - \\
\hline
Network Ops & 
Offer reliable QKD networks (\hypertarget{R16}{R16}).&
Keep networks secure.  \\
\hline
Promoters & 
Gather data for research calls; promote research calls; organise events. & 
Disseminate migration insights; create awareness; encourage ownership of quantum threat; facilitate/stimulate knowledge sharing; facilitate stakeholder mixing; encourage cryptoagility; matchmake research calls with research groups (\hypertarget{R17}{R17}). \\
\hline
Regulators & 
Set regulation (\hypertarget{R18}{R18}); create transition policy; mandate interoperability; set minimum transition requirements (\hypertarget{R19}{R19}). & 
Give and receive influence with standardisation bodies (\hypertarget{R20}{R20}); set regulation goals (\hypertarget{R21}{R21}). \\
\hline
Research \& Education & Disseminate knowledge; develop new PQC schemes; do cryptanalysis; develop talent; certify consultants (\hypertarget{R22}{R22}).
& Create awareness; interpret knowledge from fundamental research for applications (\hypertarget{R23}{R23}).\\
\hline
Standards & 
Make standards; provide advice to regulators (\hypertarget{R24}{R24}). &
Prevent fragmentation by mandating interoperability; give and receive influence from regulators (\hypertarget{R25}{R25}).
\\
\hline
Supervisors & 
Specify regulation details (\hypertarget{R26}{R26}); enforce regulation compliance (\hypertarget{R27}{R27}). &
Translate regulation towards practice (\hypertarget{R28}{R28}); influence regulators (\hypertarget{R29}{R29}).\\
 \\
\bottomrule
\end{tabular}
\end{minipage}
}
\end{table}
\subsubsection{Visibility and influence of NIS actors}\label{visibility_and_influence}
With the boundaries agreed and twelve actors validated (\cref{fig:validated_actors}), we consider the forms of power present by examining the prominence and influence of specific actors.
In this innovation system, two forms of direct influence were identified; first, the ability to finance activities.
Although Special End Users were identified as having buying power, they were not consistently viewed as sources of finance.
In fact, Financiers were the only actors identified to be financing the transition, with a consultant noting they are `putting money on the table'.
Participants also tended to equate funding with steering, as a financial services supervisor remarked that Financiers are
`not only financing but also steering'.
The second form of influence is the ability to assign accountability and liability, exercised through obligations. 
A civil servant explained that regulation can be used `to implement this accountability and responsibility,' and Regulators were therefore consistently viewed as the most influential actors.
Supervisors were acknowledged to play a role in enforcing regulation, although this influence was downplayed by Regulators and Supervisors themselves. 
A supervisor in the financial services framed supervisory influence as ad hoc and incidental, opining that
`it just so happens that these organisations also influence regulations, but it's not a common thing for supervisors'. 
Thus, while Supervisors hold significant enforcement power, they are perceived as less influential than Regulators.
The discarding of actors also revealed ambiguity about the influence of some groups, most notably big tech, whose prominence was acknowledged but not consistently tied to significant influence.
With this understanding of system composition and dynamics, we examine the responsibilities of the actors for \ref{RQ2}.
\subsection{\textit{Actor Roles and Responsibilities}}\label{arr}
The participants were sorted into three groups, and two groups were led to separate rooms.
The twelve actors were distributed among the groups, with each group discussing four actors.
\subsubsection{Articulating roles and responsibilities}
\begin{figure}
    \centering
    \includegraphics[width=0.4\linewidth]{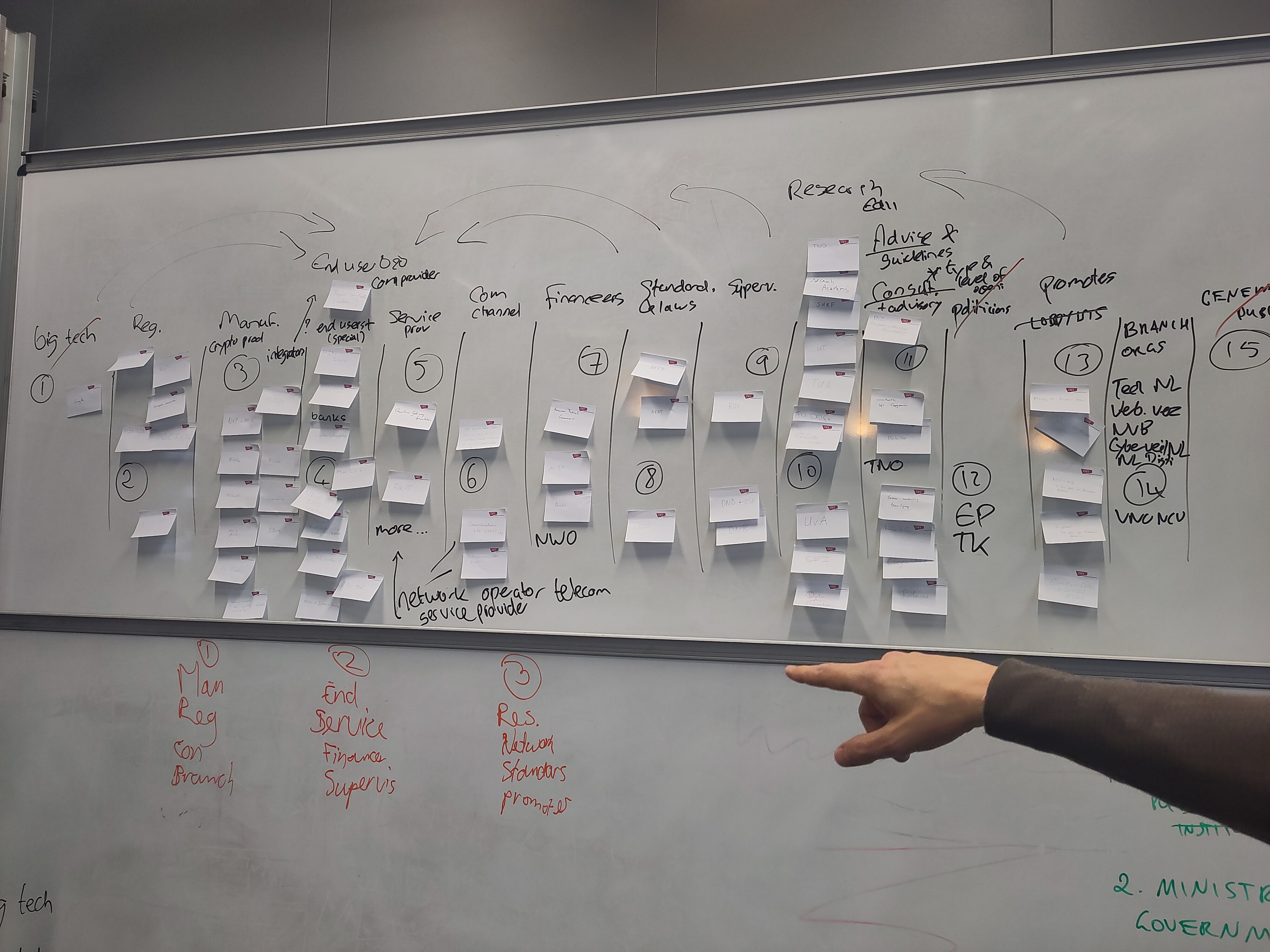}
    \caption{The final board from the Mapping exercise contained the initial 15 proposed categories.}
    \label{fig:categories}
\end{figure}

In each breakout group, the assigned actors were discussed in turn, with participants suggesting roles and responsibilities and leads noting down the roles and responsibilities that emerged.
During this process, a financial services supervisor remarked 
`when you say [the quantum threat] should be embedded in the risk management, for me that's actually already the next phase.
First you need to be aware there is a threat.'
This delineation of \textit{formal} roles, such as including the quantum threat in a risk assessment, and \textit{interpretative} roles, such as recognising the relevance of the quantum threat, is instructive in our analysis.
We describe the formal roles in \cref{formal_roles}, the interpretative roles in \cref{soft_roles} and reflect on ambiguities and gaps identified in \cref{gaps}.
Formal and interpretative roles and responsibilities are listed in respective columns of \cref{tab:roles_responsibilities_from_transcripts}.
\textit{\cref{fig:categories} about here.}
\subsubsection{Formal roles and operational responsibilities}\label{formal_roles}
Financiers and Promoters were the two actors identified with funding roles. 
Financiers directly support fundamental research, applied research, and startups, while Promoters contribute indirectly by collecting data for research calls, promoting them, and organising related matchmaking activities.
In development roles, Research \& Education institutions were seen as central to knowledge creation and dissemination, developing new PQC schemes, and performing cryptanalysis to improve scheme maturity. 
It was suggested that they may play a future role in validating QSC expertise of Consulting \& Advisory organisations. 
Network Operators contribute by building QKD networks, and Manufacturers of Cryptography Components by integrating PQC into their products. 
Migration Service Providers develop and maintain scalable migration services.
Migration Service Providers also support implementation activities by offering standard tooling and bespoke migration services, while Consulting \& Advisory organisations can implement QSC for End Users.
For regulation, Regulators set and enforce regulation and develop policy. Standardisation Bodies were viewed as central to creating standards, supported by Branch organisations that define best practices. 
Supervisors and Standardisation Bodies share responsibilities for standardisation, particularly in formulating and enforcing regulation within the \hyperlink{nisqscn}{NISQSCN}. 
As one supervisor explained:  
`They will set a goal, and they don't know how the goal will be reached, and we write the regulations, this is the way that you can actually fulfil your goal, and they then decide, yes this sounds like a good plan, or not' [Supervisor-1, Breakout 2].
Equally important as these formal roles are the interpretative roles that subtly shape the \hyperlink{nisqscn}{NISQSCN}, explored next.
\subsubsection{Interpretative roles and soft influence}\label{soft_roles}
Awareness creation was described as a shared responsibility which is aimed at End Users. 
Beyond disseminating knowledge, Research \& Education institutions help build wider awareness. 
End Users manage their own internal awareness by understanding the threat and initiating early migration. 
Regulators and Promoters also contribute, with regulators targeting influential figures such as CISOs to prompt early, `no-regret' actions.
On Expertise, participants discussed both technological and regulatory knowledge. 
Research \& Education institutions play a major role, acting as misinformation radars (`a bulls*** radar') for emergent news stories and `bridg[ing] the gap' between theory and industrial applications. 
End User security specialists negotiate migration details internally, while supervisors in the \hyperlink{nisqscn}{NISQSCN} `translate' broad regulation into practical requirements for their sectors.
Coordination roles were also highlighted, with Consulting \& Advisory firms balancing cooperation with competitors against commercial interests. 
End Users coordinate internally to avoid silos and externally by signalling demand for PQC-enabled services. 
Regulators identify public-interest outcomes that may diverge from commercial incentives and set regulation accordingly:  
`We want to take the societal impact... and put countervailing pressure on what the end user might decide... we write certain regulations to make sure that they are forced to care.'
A major challenge is avoiding fragmentation, described by an engineer as as the `biggest fear' in the transition, greater even than quantum attack; we will examine this issue in the following section.
Finally, several actors exert interpretative influence: Financiers shape priorities through funding decisions, Promoters stimulate development, and Supervisors, Standardisation Bodies, Regulators, Promoters and Financiers all influence system evolution through targeted financial and organisational support.
These findings show that soft influence and interpretative roles are essential in guiding the transition to QSC.
Next, we explore these responsibilities further, identifying system gaps relevant to \ref{RQ3}.
\subsection{\textit{Managing the Transition}}\label{mt}
We employ sociotechnical imaginaries in our response to \ref{RQ3}.
However, the sociotechnical imaginaries do not emerge directly from the workshop results, but rather from our subsequent interpretation of those results.
Hence, in this section we will create a foundation for answering \ref{RQ3} by setting out the relevant points from the workshop data and subsequently in \cref{sec:discussion} we will reflect on the results, define the sociotechnical imaginaries, and address \ref{RQ3}.
\subsubsection{Gaps, frictions, and misalignments}\label{gaps}

Participants identified several gaps and tensions within the \hyperlink{nisqscn}{NISQSCN}.
First, two breakout groups independently highlighted high institutional inertia around updating cryptographic systems:
`[Organisations] don't want to change anything... unless it's absolutely necessary' [Government-1, Breakout 2];
`people will not change cryptography unless they absolutely have to' [Research-4, Breakout 1].
This inertia is amplified in high-security contexts such as critical infrastructure where reliability is paramount.
Although PQC schemes have been implemented by some pioneers, all breakouts expressed doubts about the maturity or breadth of current PQC:
`this new cryptography is not very mature' [Government-1, Breakout 2];
`[research and education is] filling the gap... researching new cryptographic algorithms' [Industry-1, Breakout 1];
`manufacturers... are in a very immature stage' [Government-4, Breakout 3].
Maturity is central: as cryptographic schemes are implemented, bugs are fixed, and trust increases both implicitly through use and explicitly through cryptanalysis.
A second gap concerned reliance on open-source PQC. 
Most organisations expect to depend on tools like Open Quantum Safe\footnote{\scriptsize{\url{https://openquantumsafe.org/}}}, although technical experts stressed that no actor is formally responsible for ensuring quality of open-source tools:
`No-one is in charge, no-one is liable for it' [Research-1, Breakout 3].
Participants suggested commercial adopters had a `responsibility' to help maintain these libraries [Branch-1, Breakout-3; Advisor-1, Breakout-3], and some called for a stronger EU role to prevent malicious updates and ensure accountability:
`we fund [open-source PQC] early, we know who is in charge... and we don't have [an attacker] pushing updates to billions of devices' [Research-1, Breakout 3].
Orchestrated open-source governance was even framed as `more important than post-quantum security itself' [Research-1, Breakout-3].
Open-source development was described as serving the public good, `if Digi-Connect... makes a library safer, that benefits all citizens' [Research-1, Breakout 3], yet its current form was characterised as ad hoc: `the people in big tech using it kind of contribute... It’s a band-aid, it’s not a system' [Research-1, Breakout 3]. 
The Linux ecosystem was cited as an example of successful open-source that could be emulated for PQC.

A further tension involved End User responsibility. 
We recall from \cref{arr} that End Users have a responsibility to recognise the need to transition, with Regulators and Promoters typically communicating this responsibility. 
This was described as a `challenge', with the typical reaction to such outreach summarised by a regulator as
`\textit{yeah, it’s a big thing!} But nothing happens' [Government-4, Breakout 3].
Participants highlighted a lack of ownership:
`There has to be ownership... I don't think they all feel that responsibility.
They might think \textit{I have 10 other problems that I need to address right now}... we're still trying to make sure they own it' [Government-4, Breakout 3].
Another group similarly noted that awareness does not guarantee action:
`there is a big group that is aware but is just sitting on their hands' [Supervisor-1, Breakout 2].
More broadly, a `lack of demand' from End Users was reported.
Motivations vary: timing (`their time is not right yet... they don’t have to be pioneers' [Advisor-2, Breakout 2]), uncertainty (`why would you... invest this much money, and then different standards come out?' [Advisor-1, Breakout 3]), or perceived irrelevance (`Why would they target this small company?' [Advisor-2, Breakout 2]).
Even when the threat is acknowledged, limited tooling and immature cryptography were framed as hindering action:
`we can start screaming \textit{you should migrate}, but if you don’t have the tools... you create an extra obstacle' [Government-3, Breakout 3].
Relatedly, incentives and penalties, described by two Breakouts as `carrots and sticks', were seen as weak.
End Users have few genuine incentives:
`Okay, I can see the stick, but what’s the carrot?... The carrot is the same thing as the stick!' [Research-1, Breakout 3].
The only clear incentive is the `PR move', which is relevant to pioneers, a small minority of End Users.
Incentives were viewed as powerful levers, with obligations seen as less desirable but often effective:
`The carrot [is] always better... but sometimes you have to use the stick' [Branch-1, Breakout 3].
The stick was understood as enforcement of regulation:
`You can't chase after them with a stick while you don’t have any clear goals' [Advisor-1, Breakout 3].
The absence of mandates was therefore framed as a major barrier to transition.
Participants cited schemes such as the EU cloud certification\footnote{\scriptsize{\url{https://ec.europa.eu/newsroom/cipr/items/713799/en}}}, arguing that without government-backed certification it is `hard to chase after them' [Advisor-1, Breakout 3]. 

Finally, we return to the issue of fragmentation. 
Across multiple breakout groups, participants emphasised the danger that countries will adopt cryptographic standards that are not mutually compatible (`It's really bad if, for instance, one country says \textit{I don’t want hybrid} and another says \textit{we must have hybrid}' [Research-4, Breakout 1]); avoiding this requires standards that are sufficiently `lenient' to accept one another.
This is a substantial task: the specification of cryptographic standards involves numerous choices (e.g. scheme, parameters, security level).
Given the difficulty of changing cryptography and the many options, technical experts perceived the likelihood of natural interoperability as low; as it is extremely hard to implement retrospectively, interoperability must be intentional.
Standardisation bodies could coordinate internationally to develop technical standards which prevent fragmentation.
However, as noted by one researcher, coordinating externally to ensure leniency towards others' standards runs against the core purpose of standardisation bodies, `it attacks their autonomy', as their role is typically to narrow, not broaden, acceptable choices.
Thus, while Supervisors and Standardisation bodies may willingly coordinate at a domestic level to avoid fragmentation, accepting standards from other countries is not a neutral act; as noted in \cref{trust}, trust is constitutive of this process. 
We summarise that Regulators have the authority to mandate interoperability, but depend on Standardisation bodies and Supervisors to determine the details.
\subsubsection{Managing uncertainty, governing alignment, and navigating divergent imaginaries}
Having examined key gaps and tensions within the \hyperlink{nisqscn}{NISQSCN}, we consider how organisations should act under conditions of uncertainty. 
Although no CRQCs currently exist, and estimates for their arrival vary widely and are not guaranteed, the question of whether to migrate to quantum-safe cryptography was never raised.
Participants consistently treated migration as a necessity, frequently citing the \textit{store-now, decrypt-later} (SNDL) threat [\cite{timelines}] and historically long transition timelines: `That's why it's thirty, forty years till the new system, like AES and RSA' [Industry-1, Breakout 1].
This was summarised well by one participant expressing the view that all organisations should migrate: `it should be everyone'.
We will reflect on the implication of this alignment in \cref{sec:discussion}.

Having established in \cref{trust} that trust is paramount in cryptography innovation systems, we examine how governance can facilitate alignment and legitimacy.
First, the US requirement for public organisations to migrate by 2035 was cited often and framed positively:
`they have one big advantage: they set a clear date' [Branch-1, Breakout 3].
Setting a `dot on the horizon' was repeatedly endorsed:
`I think it helps if European Parliament sets... clear milestones.'
Similar action was anticipated at the European level, rather than national: `I think as the Netherlands we probably won’t [set migration dates]. 
We will do it in a European collaboration' [Government-4, Breakout 3].
Second, two breakout groups independently cited EU standardisation of USB-C chargers\footnote{\scriptsize{\url{https://commission.europa.eu/news-and-media/news/eu-common-charger-rules-power-all-your-devices-single-charger-2024-12-28_en}}} as an example of standardisation bodies influencing regulators to ensure interoperability. 
This was framed as prioritising decisive action over finding the perfect option:
`it’s not like they developed USB-C... it was just becoming the standard anyway' [Advisor-1, Breakout 2].
However, choices in cryptographic standardisation must be made carefully. 
While EU processes mitigate this risk by involving multiple cybersecurity agencies, it cannot be eliminated entirely. 
Thus, the development of standards, central to this innovation system, is both technical and political.
We will reflect on what this implies for standardisation and regulation in \cref{sec:discussion}.

Finally, we report points voiced by participants which relate closely to quantum-safe futures.
Participants were united in viewing migration as essential for mitigating the quantum threat and broadly agreed on what constitutes public benefit.
As one participant noted, `faster adoption [is] for social good' [Industry-1, Breakout 1]. 
Moments of disagreement revealed competing definitions of success, most clearly when Branch-1 cautioned that consultancy–provider partnerships risk creating `lock-in' and argued that branch organisations should instead define best practices [Branch-1, Breakout 3]. 
This highlights a divergence between the Consulting \& Advisory imaginary and that of other actors.
We will articulate these imaginaries and reflect on their implications for governance in \cref{sec:discussion}.
\begin{figure}
    \centering
    \includegraphics[width=0.4\linewidth]{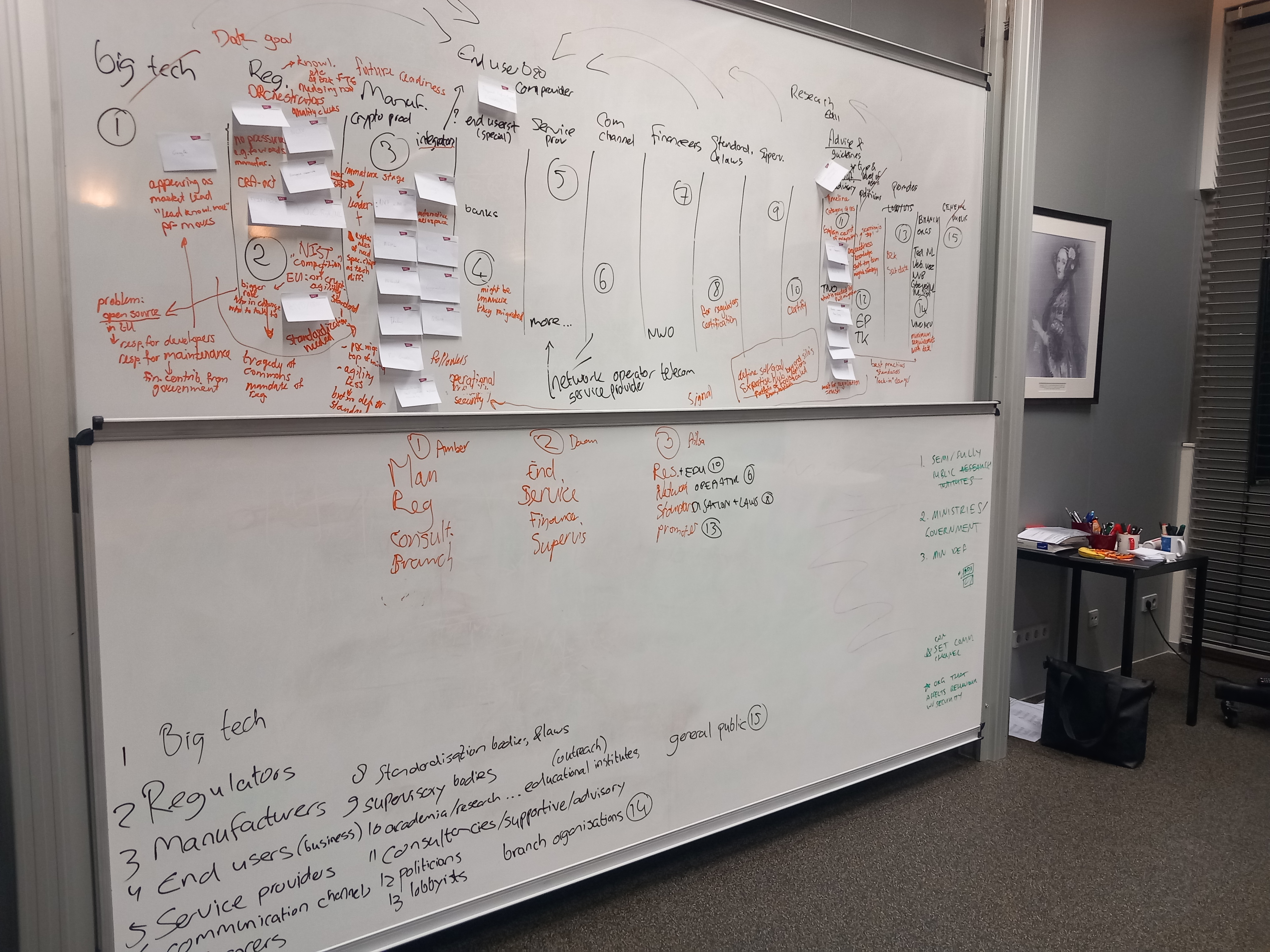}
    \caption{The lead of Breakout 3 annotated the original board in orange to record key points.}
    \label{fig:final_board}
\end{figure}
\subsection{\textit{Summary}}\label{sec:summary}
We used the workshop output to address each research question in turn. 
For \ref{RQ1}, \cref{mdis} identified the categories of actors present in the innovation system, presented in \cref{fig:validated_actors}. 
For \ref{RQ2}, \cref{arr} outlined the formal roles that emerged from the transcripts, shown in the second column of \cref{tab:roles_responsibilities_from_transcripts}, and \cref{arr} presented the organisational dynamics observed within the \hyperlink{nisqscn}{NISQSCN}, including interpretative roles, which are summarised in the third column of the table.
Finally, in \cref{mt} we set out a foundation for addressing \ref{RQ3} by presenting the gaps and tensions which were observed in this innovation system.
Participants consistently highlighted the limited availability of migration tools, and described several points of divergence relating to standards, cryptographic maturity, obligations, and milestones.
They also identified tensions between individual organisational incentives and broader system needs, and variation in expectations regarding the use of regulatory levers.

The workshop design generated additional methodological insights.
Participants identified a larger set of system actors than anticipated, which required real-time adjustments to the Breakout activity.
This adaptation enabled a more detailed mapping of the innovation system and supported a wide range of contributions.
However, we acknowledge that this adaptation required methodological flexibility and limits the generalisability of some results.
We reiterate the operational insight that technical expertise of the workshop lead ensured an accurate framing of the discussions and enabled strong stakeholder communication throughout the study.
We interpret these findings in the following section to answer \ref{RQ3}.
\section{Discussion}\label{sec:discussion}
This section discusses the findings in relation to the research questions, connecting them to innovation system literature, organisation theory and imaginaries.
\textit{\cref{fig:final_board} about here.}
\paragraph{Innovation system boundaries and dynamics (\ref{RQ1}).}
We reflect on the implications of \cref{mdis}.
Key decisions made during the scoping exercise demonstrated that the boundaries of the \hyperlink{nisqscn}{NISQSCN} are not given but negotiated, shaped by values, institutional dynamics, and perceptions of influence.
Disagreements over how to frame the innovation system and whether to include NIST, cryptoagility, or particular End User subgroups revealed that the setting of boundaries functions as a normative act as much as an analytical one, which agrees with our theoretical framing [\cite{Lundvall01022007,nelson1993nis}].
Our results also indicate that power is interpreted rather than simply possessed, as even actors with formal authority, such as Supervisors, were perceived as less influential than Regulators, which supports accounts of the politics of responsibility and system alignment [\cite{Stirling31122024, geels2002technological}].
We note a connection between \ref{RQ1} and \ref{RQ3}: system boundaries and visions of a quantum-safe future both operate as sociotechnical imaginaries which must be continually negotiated, shaping the very governance structures intended to steer transition.
\paragraph{Roles and organisational dynamics (\ref{RQ2}).}
We analyse the roles and responsibilities outlined in \cref{arr}. 
The system was mapped (\cref{fig:validated_actors}) by inspecting responsibilities in \cref{tab:roles_responsibilities_from_transcripts} which occasion some transfer between actors.
Two core processes emerged: a \textit{technological} process and a \textit{governance} process.
As noted in \cref{arr}, awareness alone does not translate into action. 
Fostering End User ownership surfaced as a distinct but nebulous responsibility, often assigned to Regulators and Promoters. 
Current incentives are insufficient to drive migration, and even when the threat is recognised, immature tooling and cryptographic uncertainty hinder progress. 
Proactive system management must therefore address the maturity and trustworthiness of PQC.
Participants emphasised the need for cryptanalysis and the importance of high-quality, well-governed open-source PQC software, which presents an opportunity for coordinated public-good–oriented development.
In practice, the technological process is shaped by the few actors with financial influence, whereas the governance process is steered primarily through standards and mandates. 
Regulators hold significant authority through their mandate-setting power but rely on the expertise of Supervisors and Standardisation Bodies to set effective directives. 
Across both processes, experts act as trusted intermediaries, translating domain knowledge to support others' roles.
Given that mandates and incentives currently fall short, fostering End User ownership may require well-calibrated regulatory mandates. 
Although Regulators expressed reluctance, mandates were widely viewed as beneficial, creating market advantages for early movers.
The need for cross-border alignment, however, introduces a key unclaimed responsibility: Standardisation Bodies must coordinate externally to avoid fragmentation.

Our findings show differentiated roles across government, industry, and research, yet accountability structures remain incomplete, which aligns with prior work on distributed responsibilities in emerging technologies [\cite{Beck07022022,kong_22}].
We note that the responsibilities of motivating End Users, introducing incentives, funding cryptanalysis and increasing trust in PQC correspond to the system functions \textit{Guidance of the Search, Market Formation, Resource Mobilisation} and \textit{Legitimation} respectively [\cite{HEKKERT2007413}].
We summarise that system development depends on managing key levers: incentives and mandates, along with expert guidance.
These insights address \ref{RQ2}, demonstrating that effective transition requires mechanisms for sharing responsibility, not only technical coordination.
\paragraph{Imaginaries and governance futures (\ref{RQ3}).}
In order to address \ref{RQ3}, we analyse our findings in \cref{mt}.
As outlined in \cref{imaginaries}, although future developments of the \hyperlink{nisqscn}{NISQSCN} remain uncertain, articulating possible futures clarifies what key levers, especially governance, must ultimately address. 
Participants treated the quantum threat as urgent despite the absence of CRQCs, illustrating how imaginaries drive present-day action and supporting broad insights on anticipatory governance [\cite{beckman2024designing,jasanoff2009dreamscapes}].
Two competing imaginaries of the future \hyperlink{nisqscn}{NISQSCN} emerged: a \textit{business-driven} imaginary and a \textit{public-good} imaginary.
In the business-driven view, Consulting \& Advisory organisations lead migration services, form exclusive partnerships with providers, and create lock-ins to their advantage.
In the public-good view, Regulators set clear mandates and timelines, Standardisation Bodies coordinate externally to prevent fragmentation, and open-source PQC development is orchestrated, potentially at EU level, to ensure trustworthy, high-quality tooling. 
Public-good imaginaries emphasised that well-judged obligations can stimulate system development: regulation was seen as enabling progress when it sets clear requirements and provides sufficient time to meet them. 
In this vision, End Users follow coherent milestones, EU-level standards avoid fragmentation, and the quantum threat is mitigated with minimal wasted resources.
By articulating these imaginaries, we identify the levers available to influential actors to steer the transition toward collective benefit rather than narrow commercial advantage.
These insights respond to \ref{RQ3} by showing how imaginaries shape governance expectations and influence both perceptions of the quantum threat and preferred transition pathways, aligning well with research showing that imagined futures shape national quantum strategies [\cite{gerhold2021imaginaries, Csenkey01092024, KONG2024101884}].
\subsection{\textit{Policy and innovation system implications.}}
The Dutch case illustrates that coordinating an innovation system requires negotiating boundaries, distributing responsibilities, and aligning visions of the future. 
Policy interventions should therefore address not only technical standards but also shared visions, coordination mechanisms, and responsibility gaps. 
Drawing on our workshop and analysis, we present five recommendations for organisations central to national QSC innovation systems. 
We identify actors best positioned to lead their implementation, but highlight that all stakeholders bear responsibility for responding to the quantum threat.
\begin{itemize}
    \item \textbf{Cryptanalysis should be incentivised.}
    This implies that Financiers and Promoters should increase funding for cryptanalytic research, for example through dedicated grants, challenge programmes, and ethical-hacker bounty schemes.
    \item \textbf{Open-source PQC software should be developed and maintained.} 
    To this end, Regulators should support sustained funding models, require transparency for publicly deployed cryptographic components, and encourage industry collaboration, as in Open Quantum Safe, to ensure high-quality, well-audited open-source PQC implementations.
    \item \textbf{QSC certifications should be developed.} 
    To this end, Research \& Education institutions and Migration Service Providers should design certification frameworks that, if achieved, guarantee that a company has adequately prepared for the quantum threat.
    \item \textbf{Fragmentation should be prevented.} 
    This implies that Regulators and Standardisation Bodies should co-ordinate across borders to ensure interoperability, creating harmonised international standards, analogous to the EU USB-C directive, for example through collaborative adoption of forthcoming IETF PQC standards.
    \item \textbf{Transition milestones should be set.}
    This implies that Regulators, Supervisors, and Standardisation Bodies should work together to develop sector-specific milestones with clear timelines and mandated transition points. 
    For example, a date may be established after which vulnerable cryptography may no longer be deployed in high-risk use cases.
\end{itemize}
\section{Conclusion}
The absence of a clear link between the quantum threat and its organisational impact remains a major impediment to transition. 
Although some pioneer organisations are acting voluntarily, our findings show that current incentives and penalties are insufficient to prompt broad transition.
Governance therefore forms a foundational enabler for widespread organisational transition.
Our analysis shows that governance must address two societally critical concerns: ensuring timely transition and preventing fragmentation across contexts. 
Regulators will require the situated expertise of Supervisors to develop sector-specific mandates that balance the time required for successful transition with the risks of avoidable delay. 
Standardisation Bodies must play a central role in maintaining coherence by working across boundaries to produce unified standards, even when this requires ceding some autonomy. 
Effective quantum transition therefore depends on coordinated governance: context-sensitive regulatory mandates paired with cross-sector standardisation.
We summarise our scholarly contributions: (i) empirically grounded insight into the roles and responsibilities required for collective transition to QSC and a system-level mapping of transition actors, (ii) a methodological reflection on the workshop's iterative focus group design, which enabled detailed discussion of the innovation system and (iii) demonstration of the application-based advantage of a qualitative study led by a technical expert, which enabled strong stakeholder dialogue.
We reflect that future research could address the limitations of this research by using comparative or longitudinal designs to examine how system boundaries and responsibilities evolve across time and national contexts, and organisational case studies could inspect the sociotechnical challenges of quantum-safe transition faced in practice.
We conclude that, in the absence of governance, transition levers are insufficient to steer the innovation system towards QSC and we highlight in the Dutch context that `de basis moet op orde zijn’ (\textit{the foundation must be sound}): without a solid governance foundation, carrots will remain irrelevant, sticks will remain absent, and the public-good quantum-safe future will never arrive.

\small
\paragraph{Acknowledgements.} 
This work was supported by the NWO as part of the QISS project\footnote{\scriptsize{\url{https://www.nwo.nl/en/projects/nwa143620002}}} which is financed by the Dutch Research Council (NWO) under Grant NWA.1436.20.002; and the Quantum Delta Netherlands growthfund program under the RQSC grant\footnote{\scriptsize{\url{https://quantumdelta.nl/news/six-projects-funded-by-open-call-from-the-centre-for-quantum-and-society}}}.
The authors express gratitude to Daan Pisa and Amber Geurts of TNO Vector and Eline de Jong of UvA for support in workshop design and execution, and contributions to the revision of this manuscript.
We also acknowledge the use of UvA's AI Chat (model gpt-4o) for refining phrasing within this manuscript.
\paragraph{Declaration.}
The authors report there are no competing interests to declare.
\paragraph{Data availability.}
As the participants of this study did not give written consent for their data to be shared publicly, supporting data is not available.
\printbibliography
\appendix
\begin{table}[hbtp]
\footnotesize
\centering
\begin{tabularx}{\textwidth}{l p{0.8\textwidth}}
\toprule
\textbf{Acronym} & \textbf{Full Phrase} \\
\midrule
CRQC & Cryptographically Relevant Quantum Computer \\
CWI & Centrum Wiskunde \& Informatica\\
IETF & Internet Engineering Task Force \\
NIS & National Innovation System \\
\hypertarget{nisqscn}{NISQSCN} & National Innovation System of Quantum-Safe Cryptography in the Netherlands \\
NIST & US National Institute for Standards in Technology \\
PQC & Post-Quantum Cryptography \\
QKD & Quantum Key Distribution \\
QSC & Quantum-Safe Cryptography \\
\bottomrule
\end{tabularx}
\caption{Table of acronyms used in this paper.}
\label{tab:acronyms}
\end{table}
\end{document}